\documentclass[preprint,showpacs,preprintnumbers,amsmath,amssymb]{revtex4}
\usepackage{graphicx, latexsym, amssymb, amsmath, color, multirow, mathrsfs, booktabs, ifpdf}
\usepackage{dcolumn}
\usepackage{bm}

\def\aA{$\alpha$-nucleus\ }
\def\AA{nucleus-nucleus\ }
\def\phe6{$^6$He+$p$\ }
\def\he6pn{$p(^6$He,$^6$Li$^*)n$\ }
\def\pn{$(p,n)$\ }

\begin{document}
\title{Neutron star cooling - a challenge to the nuclear mean field}
\author{Hoang Sy Than$^{1,2}$}
\author{Dao T. Khoa$^1$}\email{khoa@vaec.gov.vn}
\author{Nguyen Van Giai$^2$}
\affiliation{$^1$Institute for Nuclear Science and Technique, VAEC \\ 179
Hoang Quoc Viet Road, Nghia Do, Hanoi, Vietnam. \\
 $^2$ Institut de Physique Nucl\'eaire, IN2P3-CNRS/ Universit\'e Paris-Sud,
 91406 Orsay, France.}

\date{\today}

\begin{abstract}
The two recent density-dependent versions of the finite-range M3Y interaction
(CDM3Y$n$ and M3Y-P$n$) have been probed against the bulk properties of
asymmetric nuclear matter (NM) in the nonrelativistic Hartree Fock (HF)
formalism. The same HF study has also been done with the famous Skyrme (SLy4)
and Gogny (D1S and D1N) interactions which were well tested in the nuclear
structure calculations. Our HF results are compared with those given by other
many-body calculations like the Dirac-Brueckner Hartree-Fock approach or
ab-initio variational calculation using free nucleon-nucleon interaction, and by
both the nonrelativistic and relativistic mean-field studies using different
model parameters. Although the two considered density-dependent versions of the
M3Y interaction were proven to be quite realistic in the nuclear structure or
reaction studies, they give two distinct behaviors of the NM symmetry energy at
high densities, like the Asy-soft and Asy-stiff scenarios found earlier with
other mean-field interactions. As a consequence, we obtain two different
behaviors of the proton fraction in the $\beta$-equilibrium which in turn can
imply two drastically different mechanisms for the neutron star cooling. While
some preference of the Asy-stiff scenario was found based on predictions of the
latest microscopic many-body calculations or empirical NM pressure and isospin
diffusion data deduced from heavy-ion collisions, a consistent mean-field
description of nuclear structure database is more often given by some Asy-soft
type interaction like the Gogny or M3Y-P$n$ ones. Such a dilemma poses an
interesting challenge to the modern mean-field approaches.

\end{abstract}
 \maketitle

\section{Introduction}
The determination of the nuclear equation of state (EOS), which is of vital
importance for the nuclear astrophysics, has been a central object of numerous
studies of heavy ion (HI) collisions for the last two decades. If the efforts in
the early 90's were concentrated on the determination of the incompressibility
$K$ of \emph{symmetric} nuclear matter (NM), for different types of the EOS are
usually distinguished by different $K$ values \cite{Be88}, recent studies of
nuclear reactions involving unstable nuclei lying close to the neutron or proton
driplines provide us with a unique opportunity to learn about the EOS of
\emph{asymmetric} NM which has a large difference between the neutron and proton
densities \cite{Ba08}. The knowledge about the NM symmetry energy (a key
ingredient in the EOS of asymmetric NM) is vital not only in studying the
dynamics of HI collisions involving radioactive nuclei and/or their structure,
but also in studying the neutron star formation or the $r$-process of stellar
nucleosynthesis \cite{Bet90,Su94,Lat04}. In particular, the most efficient
process of the neutron star cooling, the so-called \emph{direct} Urca process in
which nucleons undergo direct beta (and inverse-beta) decays
\cite{Lat91,Lat94,Pa04}, can take place only if the proton-to-neutron ratio
exceeds 1/8 or the proton fraction $x\geq 1/9$ in the $\beta$-equilibrium. The
latter is entirely determined from the density dependence of the NM symmetry
energy $S(\rho)$ by the following balance equation \cite{Lat04}
\begin{equation}
 \hbar c (3\pi^2\rho x)^{1/3}=4S(\rho)(1-2x), \label {e1}
\end{equation}
where $\rho$ is the total NM density. It is of fundamental importance whether
the direct Urca process is possible or not. If the $x$ value cannot reach the
threshold for the direct Urca process, then the neutron star cooling should
proceed via the indirect or \emph{modified} Urca process which has a reaction
rate of $10^4\sim 10^5$ times smaller than that of the direct Urca process and
implies, therefore, a much longer duration of the cooling process
\cite{Lat91,Lat94}. Although a recent test \cite{Kl06} of the microscopic EOS
against the measured neutron star masses and flow data of HI collisions has
shown that the direct Urca process is possible in some cases if the predicted
neutron star mass is above a lower limit of $1.35\sim 1.5$ solar mass
($M_\odot$), the overall cooling time of a neutron star is still unknown as yet
\cite{Lat04} due to the uncertainty about the high-density behavior of
$S(\rho)$. We also note here a very strong influence by the neutron
superfluidity in the inner crust of neutron star \cite{Lat94,Ya00,Ya04}, where
the superfluid effects can reduce the cooling time by a factor of $3\sim 4$. In
the same direction, the cooling time might be further shortened by the possible
phase transitions to quark matter or pion condensate occurring in the core of
neutron star and leading, therefore, to a much higher central density and a
smaller star radius \cite{Lat94}.

Microscopic studies of the EOS of asymmetric NM have been performed in both
nonrelativistic and relativistic nuclear many-body theories, using realistic
two-body and three-body nucleon-nucleon (NN) forces or interaction Lagrangians
(see more details in recent reviews \cite{Ba08,Bal07}). These many-body studies
have shown the important role played by the Pauli blocking as well as
higher-order NN correlations in the G-matrix used to generate the NM binding
energy at different densities. These medium effects are considered as physics
origin of the density dependence introduced into various effective NN
interactions used presently in the nonrelativistic mean-field approaches. Among
different versions of the effective NN interaction, very popular choice is the
so-called M3Y interaction which was originally constructed by the Michigan State
University (MSU) group to reproduce the G-matrix elements of the Reid
\cite{Be77} and Paris \cite{An83} NN potentials in an oscillator basis. The use
of the original \emph{density independent} M3Y interaction in the Hartree-Fock
(HF) calculation of nuclear matter \cite{Kho93} has failed to saturate NM,
leading to a collapse at high densities. Since the HF method is the first order
of many-body calculation, some realistic density dependences have been
introduced into the original M3Y interaction \cite{Kho95} to effectively account
for the higher-order NN correlations which cause the NM saturation. During the
last decade, different density dependent versions of the M3Y interaction have
been used in the HF calculations of symmetric and asymmetric NM
\cite{Kho96,Kho97,Kho05,Kho07,Basu08,Na02,Na03,Na08}, in the mean-field studies
of nuclear structure \cite{Na02,Na03,Na08} as well as in numerous folding model
studies of the nucleon-nucleus and nucleus-nucleus scattering
\cite{Kho96,Kho97,Kho05,Kho07,Kho09}. In view of these studies, it is now highly
desirable to have a realistic version of the effective NN interaction for
consistent use in the mean-field studies of NM and finite nuclei as well as in
the nuclear reaction calculations. As an exploratory step, we perform in the
present work a systematic HF study of NM using two density-dependent versions of
the finite-range M3Y interaction, the CDM3Y$n$ interactions which have been used
mainly in the folding model studies of the nucleon-nucleus and nucleus-nucleus
scattering \cite{Kho97,Kho07,Kho09} and M3Y-P$n$ interactions which have been
carefully parametrized by Nakada \cite{Na02,Na03,Na08} for use in the mean-field
studies of nuclear structure. For comparison, the same HF study is also
performed with some realistic versions of the Gogny \cite{Be91,Ch08} and Skyrme
\cite{Ch98} interactions.

\section{Hartree Fock calculations of nuclear matter}

Like other mean-field approaches, we consider in the present HF study a
homogeneous spin-saturated NM at zero temperature which is characterized by
given values of neutron and proton densities, $\rho_n$ and $\rho_p$, or
equivalently by its total density $\rho=\rho_n+\rho_p$ and its neutron-proton
asymmetry $\delta=(\rho_n-\rho_p)/(\rho_n+\rho_p)$. With the direct ($v_{\rm
D}$) and exchange ($v_{\rm EX}$) parts of the interaction determined from the
singlet- and triplet-even (and odd) components of the central NN force, the
total NM binding energy is determined within the HF formalism as
\begin{equation}
E=E_{\rm kin}+{\frac{1}{ 2}}\sum_{k \sigma \tau} \sum_{k'\sigma '\tau '}
 [<{\bm k\sigma\tau, k'\sigma'\tau'}|v_{\rm D}|{\bm k\sigma\tau, k'\sigma'\tau'}>
+ <{\bm k\sigma\tau, k'\sigma'\tau'}|v_{\rm EX}|{\bm k'\sigma\tau,
k\sigma'\tau'}>], \label{e2} \end{equation}
where $|{\bm k\sigma\tau}>$ are the ordinary plane waves. The nuclear matter EOS
is normally classified by the NM binding energy per particle which can be
expressed as
\begin{equation}
\frac{E}{A}(\rho,\delta)=\frac{E}{A}(\rho,\delta=0)+S(\rho)\delta^2+O(\delta^4)+...
\label{e3}
\end{equation}
The NM pressure $P$ and incompressibility $K$ are then calculated as
\begin{equation} P(\rho,\delta)=\rho^2{\frac{\partial}
{{\partial\rho}}}\left[\frac{E}{A}(\rho,\delta)\right]; \
 K(\rho,\delta)=9\rho^2{\frac{\partial^2}
 {{\partial\rho^2}}}\left[\frac{E}{A}(\rho,\delta)\right]. \label {e4}
\end{equation}
The contribution of $O(\delta^4)$ and higher-order terms in Eq.~(\ref{e3}),
i.e., the deviation from the parabolic law was proven to be negligible
\cite{Kho96,Zuo99} and the most important physics quantity is, therefore, the NM
symmetry energy $S(\rho)$ which is the energy required per nucleon to change the
symmetric NM into a pure neutron matter \cite{Kho96,Zuo99,Kl06,Ba08}. The value
of $S(\rho)$ at the saturation density of symmetric NM, $\rho_0\approx 0.17$\
fm$^{-3}$, is also known as the symmetry energy coefficient $J=S(\rho_0)$ which
has been predicted by numerous many-body calculations to be around 30 MeV
\cite{Kho96,Zuo99,Bra85,Pea00}. The NM symmetry energy $S(\rho)$ is often
expanded around $\rho_0$ \cite{Tsa09} as
\begin{equation}
 S(\rho)=J+\frac{L}{3}\left(\frac{\rho-\rho_0}{\rho_0}\right) +
 \frac{K_{\rm sym}}{18}\left(\frac{\rho-\rho_0}{\rho_0}\right)^2+ ... \label{e5}
\end{equation}
where $L$ and $K_{\rm sym}$ are the slope and curvature parameters at $\rho_0$.
As discussed above, the knowledge about the density dependence of $S(\rho)$ is
of vital importance in studying the neutron star formation and has been,
therefore, a longstanding goal of many NM studies using either microscopic or
phenomenological models. The main method to probe $S(\rho)$ associated with a
given (mean-field) interaction is to test this interaction in the simulation of
HI collisions using transport and/or statistical models
\cite{Ono03,Ba08,Da02,Ch05,Ba09,She07,Sh07} or in the structure studies of
neutron-rich nuclei \cite{Na08,Ch08,Ch98,Da03,Aru04,Tod05,Pie07,Pie09,Cen09}.
Based on the constraints set by these studies using the latest experimental
data, some extrapolation is then made to draw conclusions on the low- and
high-density behavior of $S(\rho)$. We show below that such conclusions still
remain quite divergent in some cases.

In this work we study asymmetric NM at different neutron-proton asymmetries
$\delta$ using two different sets of the density-dependent M3Y interaction
named, respectively, as CDM3Y$n\ (n=3,4,6)$ \cite{Kho97} and M3Y-P$n\ (n=3,4,5)$
\cite{Na08}. Concerning the first set, the \emph{isoscalar} density dependence
of the CDM3Y3, CDM3Y4 and CDM3Y6 interactions has been parametrized \cite{Kho97}
to properly reproduce the saturation point of symmetric NM and to give $K=217$,
228 and 252 MeV, respectively, in the HF approach (\ref{e2})-(\ref{e4}). These
interactions, especially the CDM3Y6 version, have been widely tested in numerous
folding model analyses of refractive \aA and \AA scattering (see a recent review
in Ref.~\cite{Kho07r}). In the present work, the \emph{isovector} density
dependence of the CDM3Y$n$ interactions is parametrized using the procedure
developed in Ref.~\cite{Kho07}, so as to reproduce the Brueckner-Hartree-Fock
(BHF) nuclear matter results for the energy- and density-dependent nucleon
optical potential (OP) of Jeukenne, Lejeune and Mahaux (JLM) \cite{Je77}. This
isovector density dependence is then scaled by a factor $\sim 1.3$ deduced
recently from the folding model analysis \cite{Kho05,Kho07} of the \pn reactions
leading to isobaric analog states (IAS) in various targets. In contrast to the
CDM3Y$n$ set, the M3Y-P$n$ interactions have been carefully parametrized by
Nakada \cite{Na08} in terms of the finite-range M3Y interaction supplemented
with a zero-range density-dependent force, to consistently reproduce the NM
saturation properties and ground-state (g.s.) bulk properties of double-closed
shell nuclei as well as unstable nuclei close to the neutron dripline. These
latest versions of the M3Y-P$n$ interaction have not been used in the HF study
of \emph{asymmetric} NM and it is, therefore, of interest to probe them here.
For completeness, the HF calculation (\ref{e2})-(\ref{e5}) has also been done
with two other popular choices of the effective NN interaction: the Gogny (D1S,
D1N) \cite{Be91,Ch08} and Skyrme (SLy4) \cite{Ch98} forces. Among them, the
latest version of the Gogny force D1N has been shown \cite{Ch08} to reproduce
the neutron matter EOS better than the older D1S version while still giving a
good description of the bulk properties of finite nuclei.
\begin{table*}
\caption{HF results for the energy $E_0\equiv E/A(\rho_0,\delta=0)$ and
incompressibility $K$ of symmetric NM, symmetry energy coefficient $J$, slope
$L$ and curvature $K_{\rm sym}$ parameters of $S(\rho)$ evaluated at
$\rho=\rho_0$ using the CDM3Y$n$, M3Y-P$n$, Skyrme (SLy4) and Gogny (D1S, D1N)
interactions. Similar results given by the DBHF calculation using the Bonn A
interaction \cite{DBHFa} and by other mean-field studies
\cite{Vlowk,Ch05,Aru04,Tod05,Pie09} are also presented for comparison.
$K_\tau=K_{\rm sym}-6L$.} \label{t0}
\begin{tabular}{|c|c|c|c|c|c|c|c|c|} \hline
 Inter. & $\rho_0$ & $E_0$ & $K$ & $J$ & $L$ & $K_{\rm sym}$ & $K_\tau$& Ref. \\
 & (fm$^{-3}$) & (MeV) & (MeV) & (MeV) & (MeV) & (MeV) & (MeV) & \\ \hline
 CDM3Y6 & 0.17 & -15.9 & 252 & 29.8 & 62.5 & 39.0 & -336 & \cite{Kho97,Kho07} \\
 CDM3Y4 & 0.17 & -15.9 & 228 & 29.0 & 62.9 & 49.8 & -328 & \cite{Kho97}\\
 CDM3Y3 & 0.17 & -15.9 & 217 & 29.0 & 62.5 & 46.2 & -329 & \cite{Kho97}\\
 M3Y-P3& 0.16 & -16.5 & 245 & 31.0 & 28.3 & -213 & -383 & \cite{Na08}\\
 M3Y-P4& 0.16 & -16.1  & 234 & 30.0 & 21.1 & -234 & -361 & \cite{Na08}\\
 M3Y-P5& 0.16 & -16.1 & 235 & 30.9 & 27.9 & -217 & -384 & \cite{Na08} \\
 D1S & 0.16 & -16.0 & 203 & 31.9 & 23.7 & -248 & -390 & \cite{Be91}\\
 D1N & 0.16 & -16.0 & 221 & 30.1 & 32.4 & -182 & -376 &\cite{Ch08}\\
 SLy4 & 0.16 & -16.0 & 230 & 32.1 & 46.0 & -120 & -396 &\cite{Ch98}\\ \hline
 DBHF & 0.18 & -16.1 & 230 & 34.3 & 70.1 & 6.88 & -414 & \cite{DBHFa}\\
 V$_{lowk}$+CT& 0.16 & -16.0 & 258 & 33.4 & 86.8 & -44.6 & -565 & \cite{Vlowk}\\
 MDI (x=-1)& 0.16 & -16.0 & 211 & 31.6 & 107 & 94.1 & -550 & \cite{Ch05}\\
 MDI (x=1) & 0.16 & -16.0 & 211 & 30.6 & 16.4 & -270 & -369 & \cite{Ch05}\\
 G2 & 0.15 & -16.1 & 215 & 36.4 & 100.7 & -7.5 & -612 & \cite{Aru04} \\
 FSUGold & 0.15 & -16.3 & 230 & 32.6 & 60.5 & -51.3 & -414 & \cite{Tod05}\\
 Hybrid & 0.15 & -16.2 & 230 & 37.3 & 119 & 111 & -603 & \cite{Pie09} \\ \hline
\end{tabular}
\end{table*}

\begin{figure}[bht] 
\includegraphics[width=0.7\textwidth]{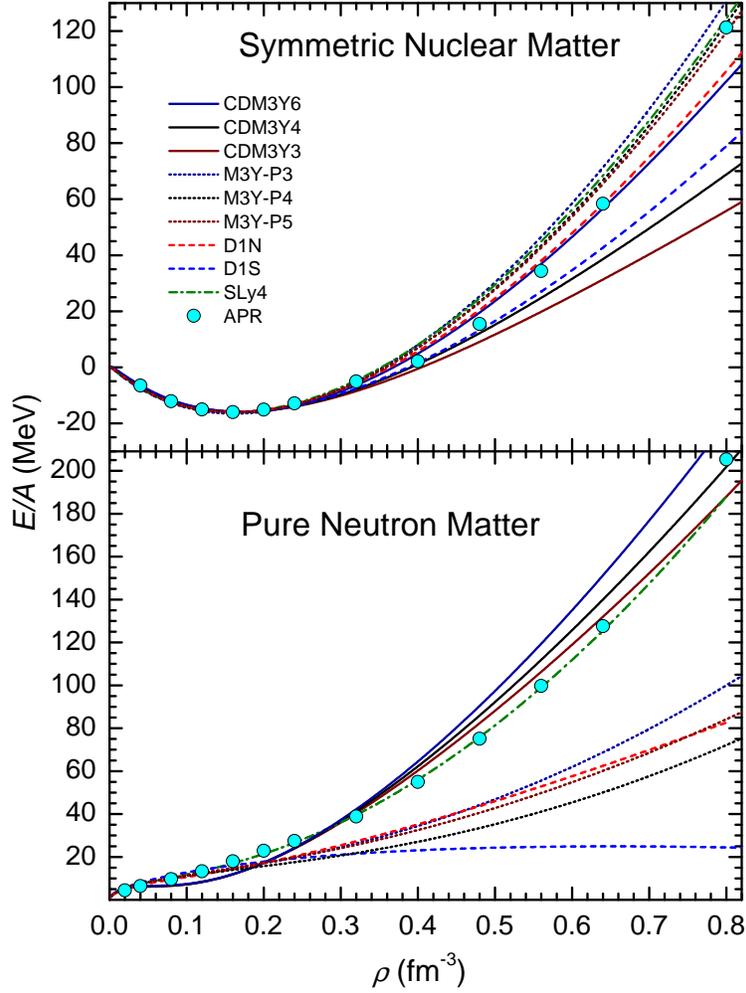}\vspace*{-1cm}
 \caption{(Color online) Energy per particle $E/A$ of symmetric NM and
pure neutron matter calculated in the HF approximation (\ref{e2})-(\ref{e3})
using the effective NN interactions given in Table \ref{t0}. Circles are
microscopic results of the ab-initio variational calculation \cite{Ak98} by
Akmal, Pandharipande and Ravenhall (APR).} \label{f1}
\end{figure}
The properties of symmetric and asymmetric NM described in the HF approximation
using CDM3Y$n$ and M3Y-P$n$ interactions are summarized in Table \ref{t0}. Since
the isovector component of the interaction does not contribute to the total
energy of symmetric NM, the interactions having about the same values of nuclear
incompressibility $K$ are expected to give similar EOS up to moderate values of
the NM density. One can see from the upper panel of Fig.~\ref{f1} that the HF
results for the energy of symmetric NM obtained with the considered
interactions, which give $K\approx 200\sim 250$ MeV (see Table~\ref{t0}), are
quite close to each other at densities up to around $2\rho_0$. At densities
above $2\rho_0$ the NM energies calculated with the M3Y-P$n$ interactions are
significantly larger than those obtained with the CDM3Y$n$ interactions, even
though the predicted incompressibilities $K$ are rather close. The main
difference here is that the M3Y-P$n$ interactions have been carefully
parametrized \cite{Na08} not only to reproduce the saturation properties of
symmetric NM like the parameter choice for the CDM3Y$n$ interactions
\cite{Kho97}, but also to give good description of the g.s. shell structure of
the magic nuclei. For a comparison, we have also plotted in Fig.~\ref{f1} the
microscopic prediction by Akmal, Pandharipande and Ravenhall (APR) \cite{Ak98}
for both symmetric NM and pure neutron matter based on the variational chain
summation method, using A18+$\delta v$+UIX* version of the Argonne NN
interaction.

For the symmetric NM, the HF results given by all considered interactions agree
well with the APR predictions at  NM densities up to $2\rho_0$, but at higher
densities the APR results seem to be closer to the CDM3Y6 and D1N curves. For
the neutron matter the picture is quite different, with the HF results given by
the CDM3Y$n$ and Skyrme (SLy4) interactions agreeing with the APR results at the
NM densities up to $4\rho_0$, while the neutron matter energies given by the
M3Y-P$n$ and Gogny (D1S, D1N) interactions are by a factor of $2\sim 3$ smaller
than the APR results in the same density range (see lower panel of
Fig.~\ref{f1}). It is obvious that such a large difference seen in the HF
results for the neutron matter energy is due to the difference in the
\emph{isovector} parts of the considered interactions. Since the isovector
density dependence of the CDM3Y$n$ interactions has been parametrized
\cite{Kho07} to reproduce simultaneously the BHF results for the isospin- and
density dependent nucleon OP by the JLM group \cite{Je77} and charge exchange
\pn data for the IAS excitations, the high-density behavior of the neutron
matter energy given by the CDM3Y$n$ interactions should approximate that given
by a BHF calculation of the neutron matter. In this sense, a similarity between
the HF results given by the CDM3Y$n$ interactions and microscopic APR results
\cite{Ak98} is not unexpected. In contrast to the CDM3Y$n$ interactions, the
isovector density dependence of the M3Y-P$n$, D1S and D1N interactions were
carefully fine tuned against the structure data observed for the neutron (and
proton-) dripline nuclei and it is also natural to expect that the EOS of the
neutron matter predicted by these interactions should be quite realistic.

\begin{figure}[bht] 
\includegraphics[width=0.7\textwidth]{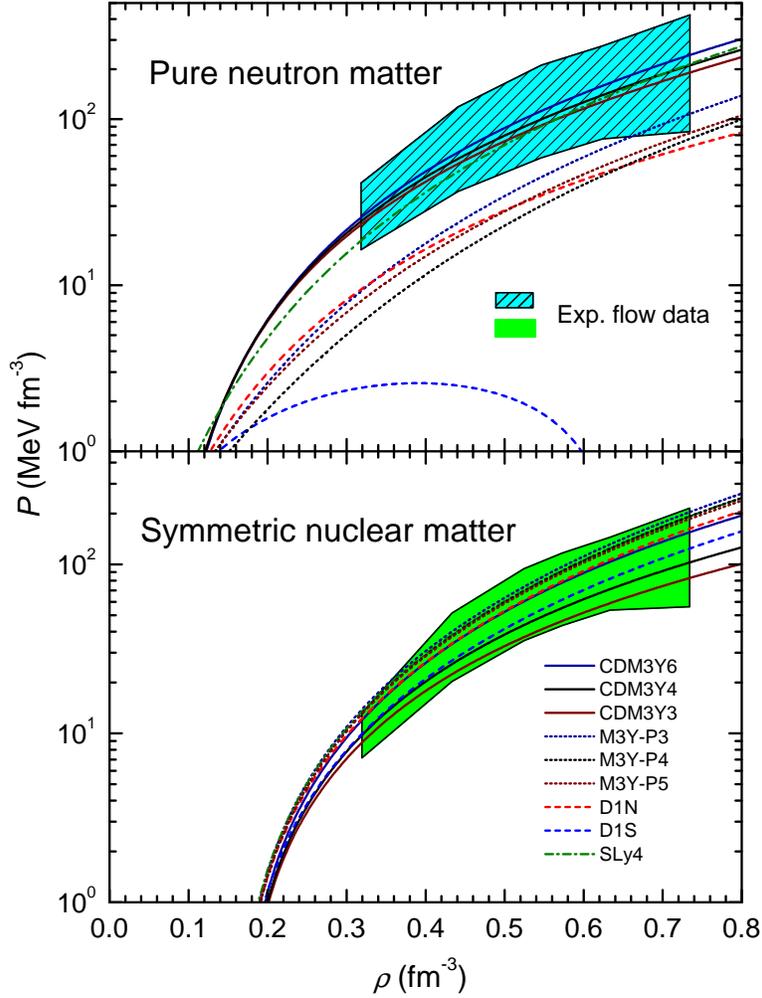}\vspace*{-1cm}
\caption{(Color online) Pressure of pure neutron matter (upper panel) and
symmetric NM (lower panel) calculated in the HF approximation
(\ref{e2})-(\ref{e4}) using the effective NN interactions given in Table
\ref{t0}. The shaded areas are the empirical constraints deduced from the HI
flow data \cite{Da02}.} \label{f2}
\end{figure}
The NM pressure (\ref{e4}) is straightforwardly evaluated from the NM energy.
The HF results for the NM pressure given by the present mean-field interactions
are compared in Fig.~\ref{f2} with the empirical constraints deduced from the
analysis of the collective flow data measured in relativistic HI collisions
\cite{Da02}. For the symmetric NM, all considered interactions give consistently
the NM pressure well within the borders of empirical data at densities up to
around $4\rho_0$ (see lower panel of Fig.~\ref{f2}). For the pure neutron
matter, the HF results given by the CDM3Y$n$ and SLy4 interactions agree nicely
with the data, while those given by the M3Y-P$n$ and Gogny interactions are
significantly below the data (see upper panel of Fig.~\ref{f2}). Among the two
Gogny forces, the D1S version gives a too low pressure in the neutron matter
which fails badly in the comparison with the data. This result confirms again
that the D1S interaction is not suitable for the study of asymmetric NM as found
in the previous NM studies \cite{Ak98,Fr81,Wi88}. Looking at Fig.~1 of
Ref.~\cite{Ch08} one might expect that a Gogny-type interaction giving a neutron
matter EOS steeper than that given by the D1N interaction and closer to the
Friedman-Pandharipande's curve \cite{Fr81} would improve the HF description of
the neutron matter pressure. From Eqs.~(\ref{e3}) and (\ref{e4}) one finds
easily that the difference observed in the upper part of Fig.~\ref{f2} is
directly related to different density dependences of the NM symmetry energy
$S(\rho)$, which in turn is determined by the isovector component of the
considered mean-field interaction.

\begin{figure}[bht] 
\includegraphics[width=0.7\textwidth]{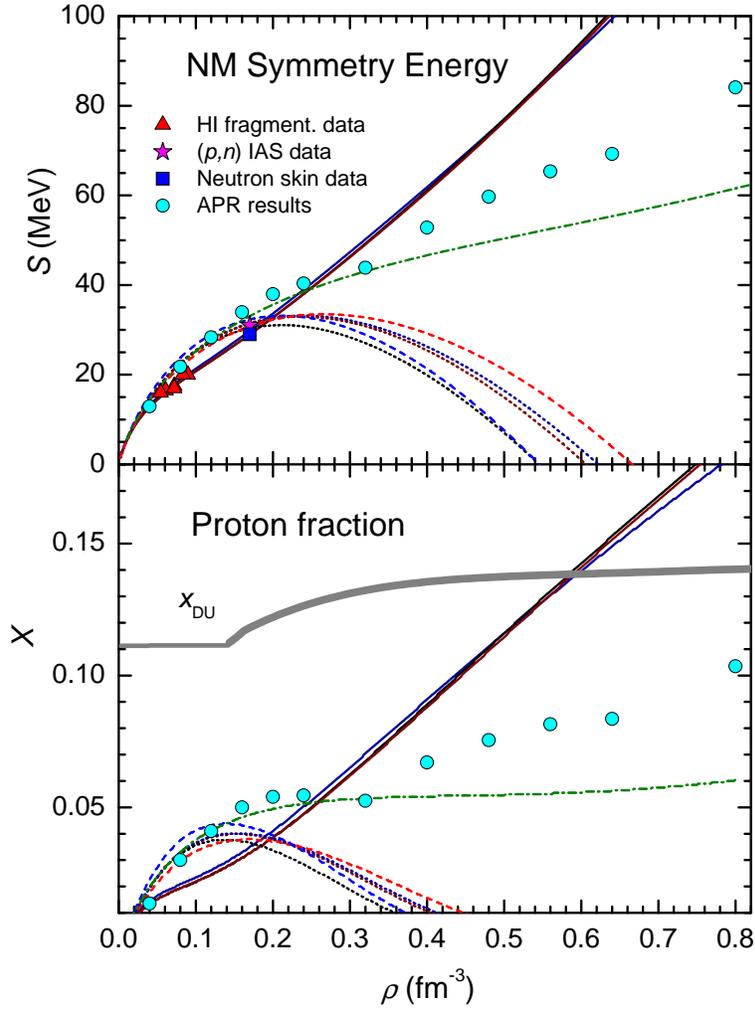}\vspace*{-1cm}
\caption{(Color online) Upper panel: NM symmetry energies $S(\rho)$ calculated
with the interactions of Table \ref{t0}. Empirical data are taken from the studies of neutron skin
\cite{Fur02}, HI fragmentation \cite{She07,Sh07,Ono03} and \pn excitation of IAS
states \cite{Kho05,Kho07}; Lower panel: Proton fractions (\ref{e1}) corresponding to these interactions,
 in comparison with the threshold $x_{\rm DU}$ for the
direct Urca process \cite{Kl06}. The curves have the same notations as in
Figs.~\ref{f1} and \ref{f2}.} \label{f3}
\end{figure}

To show further the difference caused by the isovector component of the
interaction, we have plotted in the upper panel of Fig.~\ref{f3} the HF results
for the NM symmetry energy $S(\rho)$ given by the considered interactions in
comparison with both the empirical data at low densities and microscopic APR
results. At the saturation density $\rho_0$ of symmetric NM all the models
predict the symmetry coefficient $J\approx 29\pm 3$ MeV which agrees reasonably well
with the empirical values deduced recently from the CC analysis of the
charge-exchange \pn reaction exciting the IAS states \cite{Kho05,Kho07} and
structure study of the neutron skin in medium and heavy nuclei
\cite{Br00,Fur02}.
In the low-density region ($\rho\approx 0.3 \sim 0.6\ \rho_0$) there exist some
empirical points extracted from the HI fragmentation data analysis
\cite{She07,Sh07,Ono03} and the $S(\rho)$ values given by the CDM3Y$n$
interactions are in a very good agreement with these data. The low-density
$S(\rho)$ values given by other effective interactions are slightly larger than
the data but rather close to the microscopic APR result \cite{Ak98}. Although
the studies of HI fragmentation data and/or neutron skin thickness have put some
constraints on the NM symmetry energy $S(\rho)$ at $\rho\leq \rho_0$, its
behavior at higher NM densities still remains uncertain. In contrast to the
CDM3Y$n$, SLy4 and APR predictions, the NM symmetry energies given by the
remaining mean-field interactions reach their maximal values at NM densities
around $1.5\rho_0$ and smoothly decrease to negative values at densities
approaching $4\rho_0$ (see upper panel of Fig.~\ref{f3}). These two different
behaviors have been observed earlier \cite{Ba08,She07,Ba05,Sto03} and are often
discussed in the literature as the \emph{Asy-stiff} (with symmetry energy
steadily increasing with density) and \emph{Asy-soft} (with symmetry energy
reaching saturation and then decreasing to negative values) behaviors. While
Sly4 and other Sly$n$ versions of the Skyrme interaction could also be assigned
to be the Asy-stiff type \cite{Ba08,Ch98,Sto03}, the Gogny and M3Y-P$n$
interactions belong definitely to the Asy-soft type.

If we compare these two behaviors in terms of the NM pressure shown in
Fig.~\ref{f2}, then the Gogny and M3Y-P$n$ interactions clearly fail to
reproduce the empirical pressure $P(\rho)$ of the pure neutron matter deduced
from the HI flow data \cite{Da02}. While this comparison might allow us to
conclude that the (Asy-stiff) CDM3Y$n$ interactions have a more appropriate
isovector density dependence compared with the (Asy-soft) Gogny and M3Y-P$n$
interactions, it is not possible to do so based on the nuclear structure results
given by the CDM3Y$n$ interactions. Namely, we have performed spherical HF
calculations for finite nuclei and found that the CDM3Y$n$ interactions give a
much worse description of the g.s. properties of light- and medium mass nuclei
with neutron excess compared with the Gogny or M3Y-P$n$ interactions. A widely
adopted procedure so far is to assume that the NM properties predicted by an
effective NN interaction should be quite realistic if this interaction gives
systematically good description of different structure properties of finite
nuclei, especially, the unstable nuclei near the neutron dripline. In this
sense, the failure of the Gogny and M3Y-P$n$ interactions in reproducing the
empirical NM pressure as well as the unsuccessful use of the CDM3Y$n$
interactions in the nuclear structure calculation pose a serious dilemma, which
makes it difficult to conclude unambiguously about the high-density behavior of
the NM symmetry energy $S(\rho)$ from the present HF results. Although there is
no accurate systematics available for all existing effective NN interactions on
their consistent use in the mean-field studies of both NM and finite nuclei, an
extensive mean-field study of NM by Stone {\it el al.} \cite{Sto03} using
different Skyrme interactions has shown that out of 87 considered versions of
the Skyrme interaction only 27 versions give the Asy-stiff behavior of the NM
symmetry, and all the remaining 60 versions give the Asy-soft behavior. It
remains, therefore, an interesting question whether the fitting procedure to
determine parameters of an effective NN interaction (from an optimal description
of the g.s. properties of finite nuclei and saturation properties of symmetric
NM), more likely ``bends" the NM symmetry energy $S(\rho)$ curve down to some
Asy-soft type rather than ``raises" it to some Asy-stiff type. We also note in
this connection a recent systematics by Kl\"upfel {\it el al.} \cite{Klu09} on
different phenomenological choices of parameters in the Skyrme-Hartree-Fock
model for a self-consistent description of nuclear structure and NM properties,
which shows that the range of NM properties still remains quite broad despite a
large sample of the nuclear ground-state properties used in the parameter fit.
It is, therefore, clear that the ability of nuclear structure data to constrain
the EOS for asymmetric NM on the mean-field level is still limited.

The difference found between the Asy-soft and Asy-stiff scenarios becomes
more drastic in terms of the proton fractions (\ref{e1}) shown in the lower panel of
Fig.~\ref{f3}. If the proton fraction $x = \rho_p/(\rho_p + \rho_n)$ exceeds a
critical value $x_{\rm DU}$, then the direct Urca (DU) process
\begin{equation}
 n\to p+e^- +\bar{\nu}_e,\ p+e^-\to n+\nu_e
\label{e6}
\end{equation}
becomes possible \cite{Lat91,Pa04}. Since the neutrino and antineutrino momenta
are negligible compared with those of protons, neutrons and electrons, the DU
threshold $x_{\rm DU}$ can be estimated from the momentum conservation and
charge neutrality condition for $p, n$ and $e^-$ only \cite{Lat91,Ste06}. We
have plotted in the lower panel of Fig.~\ref{f3} the \emph{averaged} DU
threshold as a function of the NM density taken from Ref.~\cite{Kl06}. At low
densities $x_{\rm DU}\approx 1/9$ as found by Lattimer \emph{et al.}
\cite{Lat91} in the muon-free approximation. At densities above $\rho_0$, the
charge neutrality is corrected by the muon presence which slightly enhances
$x_{\rm DU}$ \cite{Kl06}. From the HF results obtained with the considered
effective interactions, only the proton fractions given by the CDM3Y$n$
interactions can reach the DU threshold at moderate densities $\rho\approx 0.6$
fm$^{-3}$.

According to the microscopic APR study \cite{Ak98}, such a central density is
reachable in a neutron star having mass $M\approx 1.6 M_\odot$ which is well
above a lower limit of $1.35\sim 1.5M_\odot$ for the DU process established in
Ref.~\cite{Kl06}. The NM density $\rho\approx 0.6$ fm$^{-3}$ happens also to be
within the range of average central densities of the neutron star estimated from
a nuclear EOS with $K\approx 240$ MeV \cite{Lat91} which is quite close to $K$
values given by the CDM3Yn interactions (see Table~\ref{t0}). As a result, the
direct Urca process (\ref{e6}) is possible if one chooses the CDM3Y$n$
interactions for the in-medium NN interaction in the neutron star matter. Such a
scenario for the DU process is also favored by the HF results for the NM
pressure shown in Fig.~\ref{f2} where the CDM3Y$n$ interactions consistently
give the best description of empirical data for both the symmetric NM and pure
neutron matter. In contrast to the CDM3Y$n$ interactions, the choice of the
(Asy-soft) Gogny or M3Y-P$n$ interactions would definitely exclude the
possibility of the DU process because the corresponding proton fractions can
never reach the DU threshold as shown in Fig.~\ref{f3}. The microscopic APR
results obtained with the A18+$\delta v$+UIX* version of the Argone NN
interaction approach the muon-free threshold $x_{\rm DU}\approx 1/9$ only at
$\rho\approx 0.8$ fm$^{-3}$. Such a central density can exist only if the
neutron star mass $M\geqslant 2 M_\odot$ (see Fig.~11 of Ref.~\cite{Ak98}) and
the DU process is, therefore, very unlikely with the EOS given by the APR model.
In the case of SLy4 interaction, the numerical integration of the
Tolman-Oppenheimer-Volkov equation in Ref.~\cite{Sto03} has shown that the
central density reached in a neutron star having mass $M\approx 1.4 M_\odot$ is
only $\rho\approx 0.55$ fm$^{-3}$. Since the proton fraction $x$ given by the
Sly4 interaction is reaching the DU threshold at much a too high density of
$\rho\approx 1.4$ fm$^{-3}$ \cite{Sto03}, the DU process is also not possible
with the Sly4 interaction.

\begin{figure}[bht] 
\includegraphics[width=0.7\textwidth]{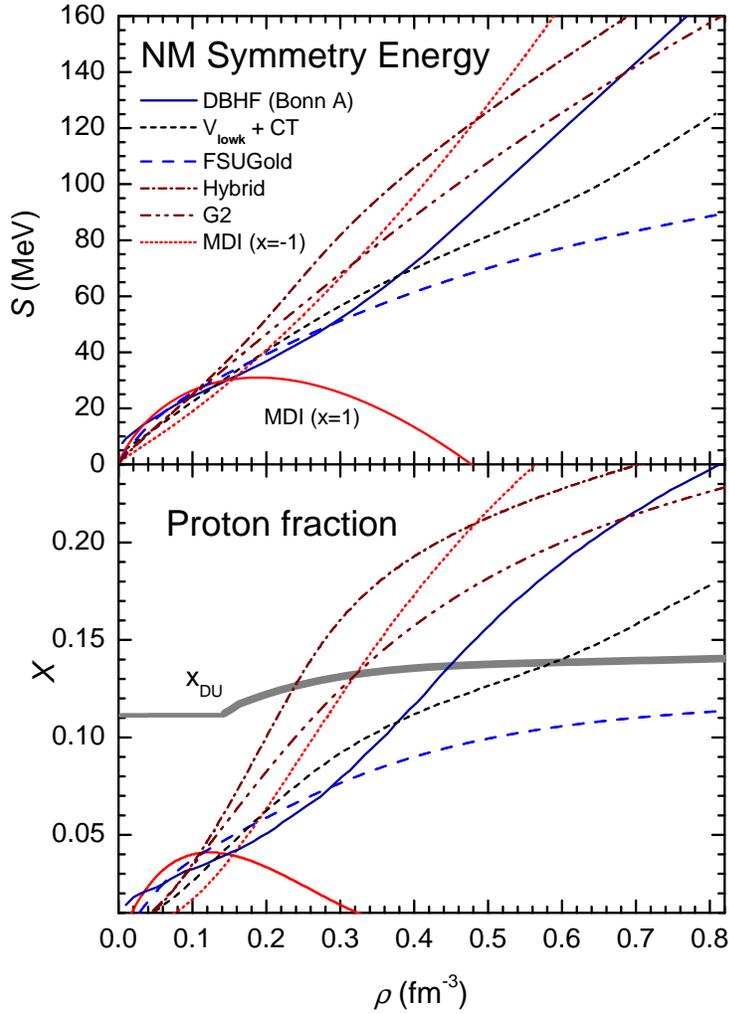}\vspace*{-1cm}
\caption{(Color online) Upper panel: NM symmetry energies $S(\rho)$ given by
different many-body studies \cite{Ch05,Aru04,Tod05,Pie09,Ak98,DBHFa,Vlowk};
Lower panel: Proton fraction (\ref{e1}) given by these symmetry energies in
comparison with the threshold $x_{\rm DU}$ for the direct Urca process
\cite{Kl06}. See more details in text.} \label{f4}
\end{figure}

The DU process has also been considered in the fully microscopic many-body
studies of the EOS using realistic free NN interactions \cite{Pa04,Bal07}, and
we found it complementary to compare the present HF results with those of a
recent Dirac-Brueckner Hartree-Fock (DBHF) study \cite{DBHFa} using an improved
treatment of the Bonn-A interaction. It can be seen from the upper panel of
Fig.~\ref{f4} that the NM symmetry energy curve given by this DBHF study is
somewhat stiffer than that given by the CDM3Y$n$ interactions. As a result, the
proton fraction estimated from the DBHF results is reaching the DU threshold
already at NM densities $\rho\approx 0.45$ fm$^{-3}$. Such a critical density
for the DU process is higher than that ($\rho\approx 0.37$ fm$^{-3}$) given by
the earlier DBHF results (see Fig.~2 of Ref.~\cite{Kl06}) and it should
correspond to a star mass above the lower limit of $1.35\sim 1.5 M_\odot$ for
the DU process. It is interesting to note that the \emph{inclusion of three-body
forces} into the many-body BHF calculations \cite{Li06} not only essentially
improves the description of the saturation properties of symmetric NM but also
\emph{gives a much stiffer NM symmetry energy} at high densities (see Fig.~4 of
Ref.~\cite{Li06}), in the opposite direction from the Asy-soft type
interactions. Given highly accurate parametrizations of the bare NN interaction,
these microscopic many-body calculations are practically parameter-free and it
is natural to assume an Asy-stiff behavior of $S(\rho)$ which allows both the
direct and indirect Urca processes to take place during the neutron star
cooling. It also is highly desirable that results of such a microscopic
many-body study can be accurately reproduced at the mean-field level using some
effective (in-medium) NN interaction which is also amenable to the nuclear
structure and/or reaction calculations. However, such ``microscopic" mean-field
interactions remain technically complicated to construct and most of the
structure and reaction studies still use different kinds of the effective NN
interaction with parameters adjusted to the optimal description of structure
and/or reaction data.

An effective NN interaction can be either fully phenomenological like the Skyrme
forces or partially based on a microscopic many-body approach like the CDM3Y$n$
interactions considered above. An interesting alternative approach has been
suggested recently by the T\"ubingen group \cite{Vlowk} which considers only the
low-momentum (below a cut-off $\Lambda=2$ fm$^{-1}$) part of the bare NN
interaction. While this ``low $k$" interaction $V_{lowk}$ still describes well
the NN scattering data up to the pion threshold, the short-range correlations
originated from the high-momentum components are treated phenomenologically at
the mean-field level. Namely, the $V_{lowk}$ has been supplemented by an
empirical density-dependent contact (CT) interaction adjusted to reproduce the
saturation properties of symmetric NM and the empirical symmetry energy $J$
within the HF approximation \cite{Vlowk}. This $V_{lowk}$+CT interaction was
shown to give also a reasonable description of the g.s. properties of some
finite nuclei including $^{208}$Pb. The NM symmetry energy and proton fraction
predicted by the $V_{lowk}$+CT interaction are shown in Fig.~\ref{f4} and they
are quite close to those predicted by the CDM3Y$n$ interactions shown in
Fig.~\ref{f3}. Like the CDM3Y$n$ interactions, the $V_{lowk}$+CT interaction
should also belong to the Asy-stiff type and allow both the direct and indirect
Urca processes during the neutron star cooling. Another famous choice of the
effective NN interaction is the Skyrme-type momentum dependent interaction (MDI)
which has been first parametrized \cite{Ch05} for the transport model simulation
of HI collisions. By varying the x parameter of the MDI interaction, the
experimental data from NSCL-MSU on the isospin diffusion have been shown to
favor the MDI (x=-1) version which gives the NM symmetry energy nearly linear in
the NM density (see Fig.~\ref{f4} or Fig.~1 of Ref.~\cite{Ch05}). One can see in
Fig.~\ref{f4} that the NM symmetry energy $S(\rho)$ given by the MDI (x=-1)
interaction is somewhat stiffer than that predicted by the DBHF calculation
using the Bonn A interaction \cite{DBHFa}. The proton fraction given by this
(Asy-stiff) MDI (x=-1) interaction is reaching the DU threshold at NM densities
$\rho\approx 0.3$ fm$^{-3}$ and is well above $x_{\rm DU}$ at the typical
central density $\rho\approx 0.5\sim 0.7$ fm$^{-3}$ of neutron star. Since the
star mass corresponding to such a central NM density should be larger than the
lower limit of $1.35\sim 1.5 M_\odot$ for the DU process \cite{Kl06}, the DU
process must be possible in this case. The MDI interaction has also been used to
describe neutron skin in finite nuclei in the Skyrme HF model \cite{Che05}, and
the MDI interaction with x between 0 and -1 was shown to reproduce reasonably
well the empirical neutron-skin data for $^{124,132}$Sn and $^{208}$Pb. However,
the situation with the MDI interaction becomes somewhat confused after the new
FOPI data on the $\pi^-/\pi^+$ ratio measured in central HI collisions at
SIS/GSI energies have been shown to clearly favor the MDI (x=1) interaction
\cite{Ba09}. In terms of symmetry energy, the MDI (x=1) interaction belongs to
the Asy-soft type (see Fig.~\ref{f4}) like the Gogny or M3Y-P$n$ interactions
considered above and it excludes, therefore, the DU process during the neutron
star cooling. Given experimental evidences favoring both the Asy-stiff and
Asy-soft versions of the MDI interaction, the behavior of the NM symmetry energy
at high densities as well as the possibility of the DU process still remain an
open question.

The NM symmetry energy has also been the subject of various relativistic mean
field (RMF) studies. In the present work, we compare our nonrelativistic HF
results with those of some recent RMF studies using carefully chosen parameters
for the energy-density functional \cite{Aru04,Tod05,Pie09}. The G2 parameter set
\cite{Aru04} has been shown to consistently reproduce the g.s. structure of
finite nuclei and bulk properties of NM. In particular, the RMF calculations
using the G2 parameters reproduce very well the empirical pressure for both the
symmetric NM and pure neutron matter \cite{Aru04}. Quite interesting is the
FSUgold parameter set developed by Piekarewicz \emph{et al.} \cite{Tod05} which
has been used to study not only the NM properties and g.s. structure of finite
nuclei but also the excitation of the nuclear giant monopole resonance (GMR).
While the FSUgold parameters give a good description of the GMR in $^{208}$Pb,
the observed GMR excitation energies in Sn isotopes could not be reproduced
using this parameter set. In order to improve the RMF description of the NM
saturation properties as well as the monopole strength distribution in Sn
isotopes, a \emph{hybrid} model of the RMF parameters has been developed
\cite{Pie09} based on the earlier NL3 model. However, the Hybrid parameters
turned out to give a worse description of the GMR in $^{208}$Pb compared with
that given by the FSUgold model and it remains, therefore, difficult to choose
between these two parameter sets. The RMF results using these parameters are
shown in Fig.~\ref{f4} and one can see that the stiffness of the NM symmetry
energy is gradually increasing as one goes from FSUgold and G2 to the Hybrid
results. It has been found by Steiner \cite{Ste06} that the RMF models typically
have a large symmetry energy and a large proton fraction, and the DU process
becomes possible at rather low NM densities. This effect can be clearly seen in
the Hybrid and G2 results shown in Fig.~\ref{f4} where the corresponding proton
fractions reach DU threshold at the NM densities of $\rho\approx 0.24$ and 0.32
fm$^{-3}$, respectively. We note further that $S(\rho)$ predicted in the Hybrid
model is very close to the RMF result by Kl\"ahn \emph{et al.} \cite{Kl06} using
the NL$\rho\sigma$ parametrization where the proton fraction is reaching the DU
threshold at $\rho\approx 0.28$ fm$^{-3}$. The behavior of the proton fraction
predicted by the FSUgold model is somewhat different from those predicted by the
Hybrid and G2 models. Namely, it approaches the muon-free threshold $x_{\rm
DU}\approx 0.11$ only at $\rho\approx 0.8$ fm$^{-3}$, like the microscopic APR
result. With the predicted maximum neutron star mass of $M\approx 1.72 M_\odot$,
the FSUgold model has been shown in Ref.~\cite{Tod05} to allow partially the DU
process in the neutron star cooling. However, if we adopt the \emph{averaged} DU
threshold taken from Ref.~\cite{Kl06} which takes into account the muon presence
at high densities, then the DU process is unlikely in this case because the
proton fraction predicted by the FSUgold model seems to saturate at densities
$\rho\geqslant 0.8$ fm$^{-3}$, around a value of $x\sim 0.11$ (see lower panel
of Fig.~\ref{f4}) like the APR results \cite{Ak98} discussed above.

Although the experimental evidences are still divergent with respect to the
Asy-stiff and Asy-soft type mean-field interactions, it is of interest to
further explore the difference between these two groups in terms of the NM
incompressibility. Using the general definition (\ref{e3})-(\ref{e4}), the NM
incompressibility $K(\rho)$ can be written explicitly in terms of the isoscalar
($K_0$) and isovector ($K_1$) parts as
\begin{equation}
 K(\rho,\delta)=K_0(\rho)+K_1(\rho)\delta^2+O(\delta^4)+...
 \label{e7}
\end{equation}
\begin{equation}
 {\rm where}\ K_0(\rho)=9\rho^2{\frac{\partial^2}
 {{\partial\rho^2}}}\left[\frac{E}{A}(\rho,\delta=0)\right];\
 K_1(\rho)=9\rho^2{\frac{\partial^2S(\rho)}
 {{\partial\rho^2}}}.
 \label{e8}
\end{equation}
It is clear from Eq.~(\ref{e8}) that the behavior of the isovector
incompressibility $K_1$ should correlate closely with the NM symmetry energy
$S(\rho)$.
\begin{figure}[bht] \vspace*{2cm}
\includegraphics[width=0.85\textwidth]{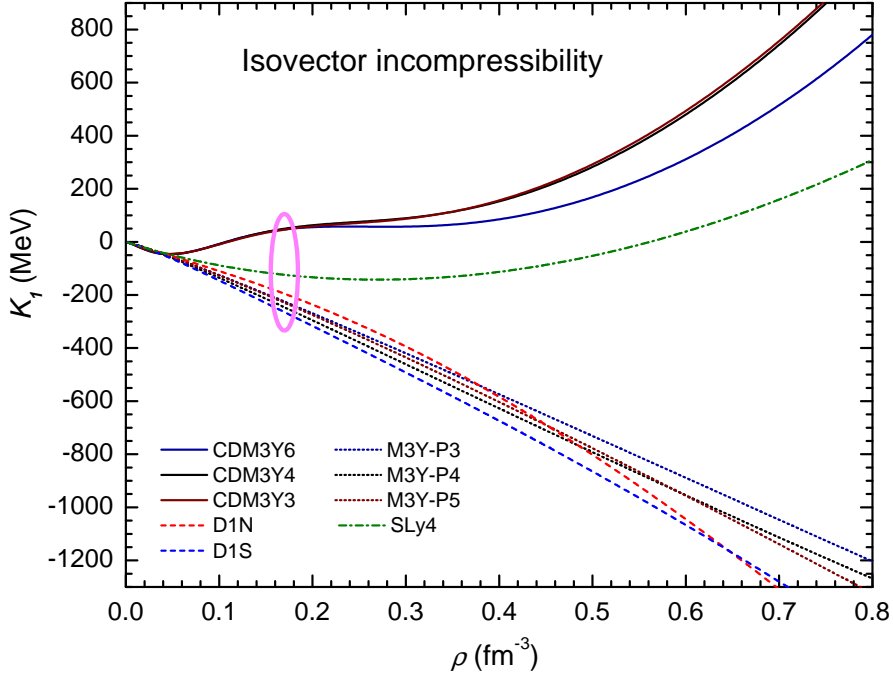}\vspace*{-3.5cm}
\caption{(Color online) Isovector part of the NM incompressibility given by HF
calculations using different effective interactions. Encircled part shows $K_1$
values near the saturation density $\rho_0$ of symmetric NM, i.e., the $K_{\rm
sym}$ values given in Table~\ref{t0}. See more details in text.} \label{f5}
\end{figure}
We have plotted in Fig.~\ref{f5} the density dependence of $K_1$ given by the
numerical differentiation of the HF results for $S(\rho)$ and one can see that
the $K_1$ value given by the (Asy-stiff) CDM3Y$n$ interactions is positive over
the density range $\rho\geqslant\rho_0$ and gradually increases to $200\sim 400$
MeV at $\rho$ approaching 0.6 fm $^{-3}$ (the DU-onset density found with these
interactions). On the contrary, the $K_1$ value given by the (Asy-soft) M3Y-P$n$
and Gogny interactions is negative over the same density range and decreases
linearly to about -1000 MeV at $\rho$ approaching 0.6 fm $^{-3}$. Since
$K(\rho)$ is actually determined from the first derivative of NM pressure
$P(\rho)$ with respect to the density, a strongly \emph{negative} isovector
incompressibility $K_1(\rho)$ should indicate a decrease of $P(\rho)$ in the
transition from symmetric NM to the pure neutron matter. Indeed, one can see in
Fig.~\ref{f2} that a decrease of $P(\rho)$ in the pure neutron matter was found
with the (Asy-soft) M3Y-P$n$ and Gogny interactions which pulls the calculated
$P(\rho)$ out of the empirical boundaries established by the HI collective flow
data \cite{Da02}. Based on this discussion, we conclude that the behavior of
isovector incompressibility $K_1(\rho)$ given by the (Asy-stiff) CDM3Y$n$
interactions is more consistent with the HI flow data compared with $K_1(\rho)$
given by the (Asy-soft) M3Y-P$n$ and Gogny interactions.

In the literature, the discussion on the isovector part of the NM
incompressibility is very often made based on the $K$ values estimated at the
saturation density $\rho_0$ of symmetric NM. It should be noted, however, that
the saturation density of asymmetric NM decreases rather quickly with the
increasing neutron-proton asymmetry $\delta$, and the pure neutron matter
($\delta=1$) becomes \emph{unbound} (see lower panel of Fig.~\ref{f1} and also
Fig.~2 of Ref.~\cite{Kho96}). As a result, $\rho_0$ is no more a stable extremum
in the NM energy curve and various expansions around it like (\ref{e5}) might
not be accurate for large neutron-proton asymmetries $\delta$. For example, the
second derivative of the approximated expression (\ref{e5}) for $S(\rho)$ gives
a purely parabolic density dependence of the isovector incompressibility,
$K_1(\rho)\approx K_{\rm sym}(\rho/\rho_0)^2$, which can strongly deviate from
the exact HF result at high densities as shown in Fig.~\ref{f5}. In addition,
the higher-order $O(\delta^4)$ term in Eq.~(\ref{e5}) has been shown by Steiner
\cite{Ste06} to be important in determining the critical density for the DU
process in neutron stars. In the vicinity of $\rho_0$, the $K_1$ values given by
the HF calculation (encircled in Fig.~\ref{f5}) are quite close to the
corresponding $K_{\rm sym}$ coefficients of the expansion (\ref{e5}), i.e.,
$K_1(\rho\to\rho_0)\approx K_{\rm sym}$. In the studies of the HI isospin
diffusion \cite{Ch05} or isospin dependence of the GMR excitation \cite{Li07},
the asymmetry of the NM incompressibility around $\rho_0$ was associated with
the quantity $K_\tau=K_{\rm sym}-6L$ which has been confined by these data to
$K_\tau\approx -550\pm 100$ MeV. The empirical $K_\tau$ value has been shown by
a recent study of the neutron-skin thickness by Centelles \emph{et al.}
\cite{Cen09} to be around $K_\tau\approx -500\pm 100$ MeV. From Table~\ref{t0}
one can see that $K_\tau$ values obtained from the (Asy-stiff) DBHF,
V$_{lowk}$+CT, Hybrid and MDI (x=-1) results are in good agreement with the
empirical values; the Asy-stiff CDM3Y$n$ interactions give $K_\tau$ values of
about -330 MeV which are somewhat above the upper limit of empirical data; the
Asy-soft MDI (x=1), M3Y-P$n$ and Gogny interactions give $K_\tau$ values of
about -400 MeV, right at the empirical upper limit. However, we note that a very
recent mean-field study of asymmetric NM using a large number of Skyrme-type
interactions by Chen {\it et al.} \cite{Che09a} has found an optimum range
$K_\tau\approx -370\pm 120$ MeV and, thus, cast some doubt on previously adopted
empirical $K_\tau$ values. In the context of our paper, this latest analysis
seems to give preference to the (Asy-stiff) CDM3Y$n$ interactions. Some further
preference for the Asy-stiff interactions can also be made based on the
empirical constraints on the $J$ and $L$ parameters established recently by the
MSU group in the analysis of the isospin diffusion and ratios of neutron and
proton spectra measured in HI collisions \cite{Tsa09}. Namely, one can see in
Table~\ref{t0} that only the $L$ and $J$ values given by the (Asy-stiff)
CDM3Y$n$ and SLy4 interactions are lying within the ranges of the double
constraint deduced from the isospin diffusion data: $L\approx 40\sim 70$ MeV and
$J\approx 30\sim 34$ MeV (see Fig.~1 of Ref.~\cite{Tsa09}). The (Asy-soft) Gogny
and M3Y-P$n$ interactions give $L$ values much lower than this limit and this
indicates again that the (Asy-stiff) CDM3Y$n$ interactions comply better with
the HI data. We note here also a similar empirical range for the slope parameter
$L\approx 45\sim 75$ MeV established recently in a systematic study of the
correlation between the neutron-skin thickness and symmetry energy \cite{War09}.
However, the neutron-skin thickness has been shown by Danielewicz \cite{Da03} to
be mainly sensitive to the surface part of the symmetry energy term in a more
elaborate mass formula for finite nuclei, while the extrapolation to the
high-density behavior of the NM symmetry energy is based more on the volume
term. Therefore, the empirical ranges for $L,\ J$ and $K_\tau$ values deduced
from the neutron-skin studies should be necessary reference points for any
mean-field study of asymmetric NM but not sufficient constraints to restrict the
behavior of the NM symmetry energy at high densities.

Finally, we note that all the mean-field calculations discussed in the present
work do not take into account the \emph{hyperon} presence in the neutron star.
The hyperon population has been estimated to make up about 18\% of the neutron
star matter and shown to significantly soften the EOS as well as reduce the
limiting neutron star mass \cite{Glen1,Glen2}. In this case, not only the direct
Urca process involving hyperons becomes possible \cite{Ya00} but also the proton
fraction is significantly enhanced by the hyperon presence. For example, the
proton fraction of a hyperon star having mass $M\approx 1.5 M_\odot$ is about
50\% larger than that of a neutron-proton-lepton star of the same mass (see
Fig.~5.28 of Ref.~\cite{Glen2}). As a result, if we assume for simplicity a 50\%
rise in the proton fractions predicted, e.g., by the microscopic APR calculation
or FSUgold model at $\rho\approx 0.6\sim 0.8$ fm$^{-3}$ (see lower panel of
Fig.~\ref{f4}), then the DU process is well allowed within these models.
Concerning a typical Asy-soft interaction like the Gogny, M3Y-P$n$ or MDI (x=1),
such a 50\% increase of the proton fraction is still not enough to make the DU
process possible.

\section{Summary}
In the framework of the self-consistent HF mean field, we have studied the bulk
nuclear matter properties predicted by two different sets (CDM3Y$n$ and
M3Y-P$n$) of the density-dependent M3Y interaction, SLy4 version of the Skyrme
interaction as well as D1S and D1N versions of the Gogny interaction. The HF
results for the NM symmetry energy and proton fraction in the
$\beta$-equilibrium are also compared with those given by the microscopic
many-body studies (DBHF and APR calculations) using the bare NN interaction, and
by the RMF studies using different parameter sets.

We have concentrated our discussion on two main aspects: the NM binding energy
and pressure in the symmetric NM and pure neutron matter, and the density
dependence of the NM symmetry energy $S(\rho)$ and the associated proton
fraction (\ref{e1}). For the symmetric NM, the main conclusion is that all the
effective NN interactions used here are more or less consistent with the
microscopic APR prediction and empirical pressure deduced from the HI collective
flow data. For the pure neutron matter, the HF predictions for the NM binding
energy and pressure show that the considered mean-field interactions are divided
into two families which are associated with two different behaviors (Asy-soft
and Asy-stiff) of the NM symmetry energy at high densities, where only the
Asy-stiff type interactions comply with the empirical NM pressure. These two
families were shown to predict two different behaviors of the proton-to-neutron
ratio in the $\beta$-equilibrium which, in turn, imply two drastically different
mechanisms for the neutron star cooling (with or without the direct Urca
process).

Although an ambiguity in the high-density behavior of the NM symmetry energy
still remains due to the experimental evidences from HI studies favoring both
the Asy-soft and Asy-stiff versions of the mean-field interaction, a comparison
of the present HF results with the empirical constraints for the symmetry
coefficient $J$ and slope parameter $L$, given by the HI isospin diffusion data
and ratios of neutron and proton spectra \cite{Tsa09} and the systematic study
of the neutron-skin thickness \cite{Cen09,War09}, seems to provide some evidence
favoring the Asy-stiff type interactions. The Asy-stiff behavior is also
predicted by the fully microscopic BHF or DBHF calculations \cite{DBHFa,Li06}
which include the higher-order many-body effects and three-body forces, and by
the latest RMF studies \cite{Aru04,Tod05,Pie09}.

A big puzzle remains why on the (nonrelativistic) mean-field level, a wide range
of the nuclear structure data can be consistently described only by using some
Asy-soft type effective interaction like the famous Gogny forces or M3Y-P$n$
interaction. In each case, the chosen parameter set for the effective NN
interaction depends strongly on the nuclear structure and/or reaction data under
consideration \cite{Sto03,Che05,Klu09,Ba09,Che09a} and there could be a plethora
of systematic uncertainties in different choices for the mean-field interaction
which are not under control and can lead, in particular, to the distinct
\emph{soft} and \emph{stiff} scenarios. In any case, the ability of available
structure and/or reaction data to constrain the EOS for high-density neutron
rich NM on the mean-field level is still limited and the most interesting
challenges are lying ahead.

\section*{Acknowledgments}
We thank Mario Centelles, Bao-An Li, Jerome Margueron, and Herbert M\"uther for
their helpful communications and discussions. The present research has been
supported by the National Foundation for Scientific and Technological
Development (NAFOSTED) under Project Nr. 103.04.07.09. H.S.T. also gratefully
acknowledges the financial support from the Asia Link Programme CN/Asia-Link 008
(94791) and the Bourse Eiffel program of the French Ministry of Foreign Affairs
during his research stays at IPN Orsay where part of the present nuclear matter
study has been performed.

\end{document}